# Decoupling between thermodynamics and dynamics during rejuvenation in colloidal glasses[*]

Xiunan Yang[1,2], Hua Tong[3], Weihua Wang[2,4], and Ke Chen[1,2,†]

[1] *Beijing National Laboratory for Condensed Matter Physics and Key Laboratory of Soft Matter Physics, Institute of Physics, Chinese Academy of Sciences, Beijing 100190, People's Republic of China*

[2] *University of Chinese Academy of Sciences, Beijing 100049, People's Republic of China*

[3] *Department of Fundamental Engineering, Institute of Industrial Science, University of Tokyo, 4-6-1 Komaba, Meguro-ku, Tokyo 153-8505, Japan*

[4] *Institute of Physics, Chinese Academy of Sciences, Beijing 100190, People's Republic of China*

We rejuvenate well-aged quasi-2D binary colloidal glasses by thermal cycling, and systematically measure both the statistical responses and particle-level structural evolutions during rejuvenation. While the moduli and boson peak are continuously rejuvenated with increasing number of cycles, the mean square displacement (MSD) fluctuates significantly between different groups of thermal cycles. The decoupling between the thermodynamical and dynamical evolutions suggests different microscopic origins for different bulk properties of glasses. We find that a small fraction of structural rearrangements triggered by thermal cycling could alter the *whole* elastic continuum and lead to the significant thermodynamic rejuvenation, while *localized* defects could be activated and deactivated at the positions close to the rearrangements with significantly high mobility change and hence result in the fluctuated dynamics even with only about 10% of particles as fast regions. Our results offer a comprehensive picture for the microscopic mechanisms underlying bulk glass rejuvenation, which could be readily used to refine glass properties or to formulate further statistical theories in glassy systems.

## 1. Introduction

As a typical non-equilibrium system, glasses relax or age. During the physical aging of glassy systems, a continuous evolution of thermodynamic states towards equilibrium is coupled with the gradual slowdown of dynamics [1-5]. Intriguingly, external stimulations such as thermal cycling and mechanical stimuli are found to partially recover or "rejuvenate" the aged glass properties in bulk glasses such as metallic glasses and polymer glasses [6-12], in colloidal glasses [13-17], in spin glasses [18], and in computer simulations [19-22]. The studies of glass rejuvenation open a door for practically tuning the bulk properties of glasses and for understanding the physics underlying different glass properties. However, the microscopic mechanisms underlying rejuvenation remain unclear and a number of anomalies are hard to explain. For example, while mechanical deformations could lead to drastic increase of the relaxation dynamics of polymer glasses, some parameters concerning thermodynamic states such as equilibration time and volume recovery show no detectable change [7]; a continued thermal cycling leads to an out-sync evolution among the reversible heat of relaxation, saturated microhardness reduction, and the monotonically strengthened plasticity in metallic glasses [6].

[*] Project supported by the MOST 973 Program (No. 2015CB856800). K. C. also acknowledges the support from the NSFC (No. 11474327).
[†] Corresponding author. E-mail: kechen@iphy.ac.cn

Rejuvenation is related to a general question how glasses respond to external stimulations. Unlike the uniform atomic environment in a single crystal, each atom in glasses has a unique local structure and responds differently even to uniform perturbations. Thus, it requires the particle-level information as well as macroscopic measurements of statistical properties to understand the mechanisms underlying rejuvenation. Conventionally, bulk glass experiments measure macroscopic properties without particle-level details, whereas local measurements of colloidal particle motions do not directly access the macroscopic properties of thermodynamic states. However, recently-developed covariance matrix method enables researchers to investigate the bulk thermodynamic properties such as boson peak and elastic moduli in colloidal systems [23-28]. With thermal cycling and the covariance matrix method, we measure evolutions of both the local particle motions and statistical properties (moduli, boson peak, and MSD plateau) in quasi-2D binary colloidal glasses during thermal cycling. Our results demonstrate a decoupling of the evolution of thermodynamics and that of dynamics during rejuvenation. While both evolutions are induced by local rearrangements, the decoupling suggests that different structural signatures dominate different macroscopic perspectives of glasses. We find that the stiffness undermined by structural rearrangements mainly contributes to the increased plasticity and the intensified boson peak; on the other hand, the activation of localized CRRs originating from emerging high local structural entropy dominates the fluctuated MSD. Our results unveil the particle-level mechanisms underlying the evolutions of the thermodynamic and dynamic properties during rejuvenation in glassy systems.

## 2. Experimental procedure

### 2.1. Sample preparation
Two types of thermo-sensitive poly-N-isopropylacrylamide (PNIPAM) particles with different diameters were mixed to prepare colloidal glass samples [29,30]. The particle diameters were measured to be 1.1 and 1.4 μm at 22 ℃ by dynamical light scattering. A 1:1 binary mixture of the two types of particles was sufficient to avoid crystallization. A dense monolayer of the mixture between two coverslips were hermetically sealed using optical glue (Norland 65) to form quasi-2D colloidal glasses. The samples were well aged on the microscope stage for 3 h before further stimulations. There were ~3100 particles within the image frame. We used standard bright-field microscopy at 60 fps for the duration of the experiment. The trajectory of each particle was extracted using particle tracking techniques [31, 32].

## 2.2. Thermal cycling

A small amount of non-fluorescent dye (Chromatch-Chromatint black 2232 liquid, 0.2% by volume) was mixed with the colloidal suspension to absorb incident mercury light and to increase the temperature of the samples by about 0.2 K after the mercury lamp was turned on [33-35]. Before the optical heating, the packing fraction was about 0.88 and an increase of 0.2 K slightly change the packing fraction to 0.87, which is still above the jamming packing fraction (~0.85). The sample reached new thermal equilibrium in less than 1 second after the mercury lamp was turned on, and quickly recover to original temperature with original packing fraction when the mercury lamp was turned off. Thermal cycling was achieved by turning on and off the mercury lamp cyclically, similar to the non-destructive thermal cycling in bulk experiments [6]. After several trials, we found that a sequence of short thermal cycles was more efficient than a single cycle to rejuvenate the aged samples. Thus we used a group of 10 successive short thermal cycles to stimulate the colloidal glasses and measure the evolution of properties before and after the group of cycles as in Figure 1. Each cycle lasted for 40 s with 20 s light on and 20 s light off. To ensure a stable configuration without rearrangements, the data were obtained 400 s after a group of cycles. Another 9 groups of such thermal cycles were performed for the same sample and the data in the stable configuration after each group were recorded. The realization of glass rejuvenation requires proper stimulation amplitudes and well-aged initial states [15, 19], and can be technically complicated. Here we focus on the microscopic mechanisms of the rejuvenation, rather than best conditions for rejuvenation.

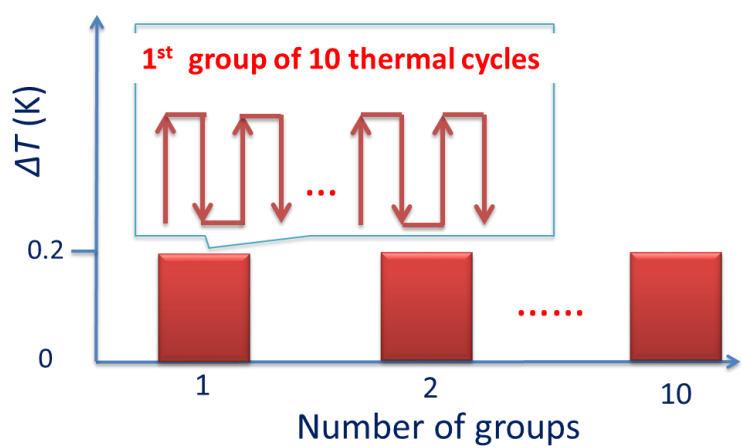

**Figure 1** 10 groups of thermal cycles were performed. Within each group, there were 10 successive short thermal cycles. For each cycle, we turned on the mercury lamp for 20 s to increase the temperature by about 0.2 K and turned off the mercury lamp for 20 s to recover the temperature to its original value. Between two groups of thermal cycles, the samples were relaxed for 400 s to ensure a stable configuration without rearrangements followed by the data measurements in 200 s.

## 2.3. Covariance matrix

The boson peak and elastic moduli before and after each group of thermal cycles are calculated based on displacement covariance matrix methods [24-27]. We first subtract the x, y positions of $N$ particles with respect to their average positions to obtain $2N$ vectors of displacements $u_i(t)$ and extract the displacement covariance matrix $C_{ij} = <u_i(t)u_j(t)>_t$, where $i,j$ run over all the $2N$ degrees of freedom and $<.>_t$ is the time average. A shadow system of particles with the same configuration and the same interaction as our colloidal system is introduced. Under the harmonic approximation, the stiffness matrix $K_{ij}$ and dynamical matrix $D_{ij}$ of the shadow system are then obtained from the covariance matrix as

$$K_{ij} = k_{BT}C_{ij}^{-1}, \ D_{ij} = K_{ij}/\sqrt{m_i m_j} = k_{BT}C_{ij}^{-1}/\sqrt{m_i m_j},$$

where $m_i$ is the mass of particle $i$. The eigenvectors of $D_{ij}$ correspond to the vibrational modes of the shadow system. We extract the boson peak intensity from the distribution of eigenfrequencies $\omega_n$ (for n = 1, …, 2N) [24]; the polarization vectors $P_n$ is employed to obtain the transverse and longitudinal spectral functions. The maxima of the spectral functions correspond to the phonon wave vector with magnitude $q(\omega_n)$ that constitute the dispersion relation. From the dispersion relation, we extract the longitudinal and transverse sound velocities, and then the shear (G) and bulk (B) modulus [27].

## 3. Experimental Results

### 3.1. Thermodynamic and mechanical rejuvenation

Figure 2a shows the distribution of eigenfrequencies $D(\omega)$ (rescaled by $\omega^{d-1}$, which corresponds to the Debye behavior in crystals) for samples from different groups of thermal cycles, where dimension $d = 2$ in our system. For each curve, an excess vibrational density of states compared to Debye model is observed as a boson peak [24]. We extract the maximum of the $D(\omega)/\omega$ as the boson peak intensity $I^*$. As in figure 2b, $I^*$ tends to increase with the number of thermal groups with few exceptions. This tendency is contrary to the decaying boson peak intensity during aging [36] and is suggestive of the occurrence of thermodynamic rejuvenation from thermal cycles. After significant rejuvenation, the samples are at high energy states and have the possibility to age again, which causes the small drop of $I^*$ after 6 groups of thermal cycles.

Figure 2c plots how bulk modulus (B) and shear modulus (G) evolve with 10 groups of thermal cycling. Moduli, especially the bulk modulus, show a roughly monotonic decrease. This decrease suggests that thermal cycling tends to continuously soften the sample with gradually refined plasticity. After 10 groups of thermal cycles, the shear (bulk) modulus decreases by more than 40% (30%). This tendency is the reversal of hardening during aging [17, 20, 37-39] and is a signature of mechanical rejuvenation. The mechanical moduli also reflect the energy states of the sample and negatively correlate with system energy [22]. The decline in the moduli thus indicates elevated energy states in our systems which is a direct evidence of thermodynamic rejuvenation in the framework of energy landscapes [19].

The coincidence of boson peak intensity increase and moduli decline during thermal cycling indicates a possible correlation between boson peak and moduli. Following previous simulations of Shintani and Tanaka [40], we fit I* and B(G) with a simple inverse relation in figure 2d. The coefficient of determination $R^2$ for the linear fitting is 0.836. This high value of $R^2$ confirms the simple relationship proposed by simulation and suggests that the boson peak intensity variation is nontrivially correlated with the change of elastic continuum characterized by the moduli decrease.

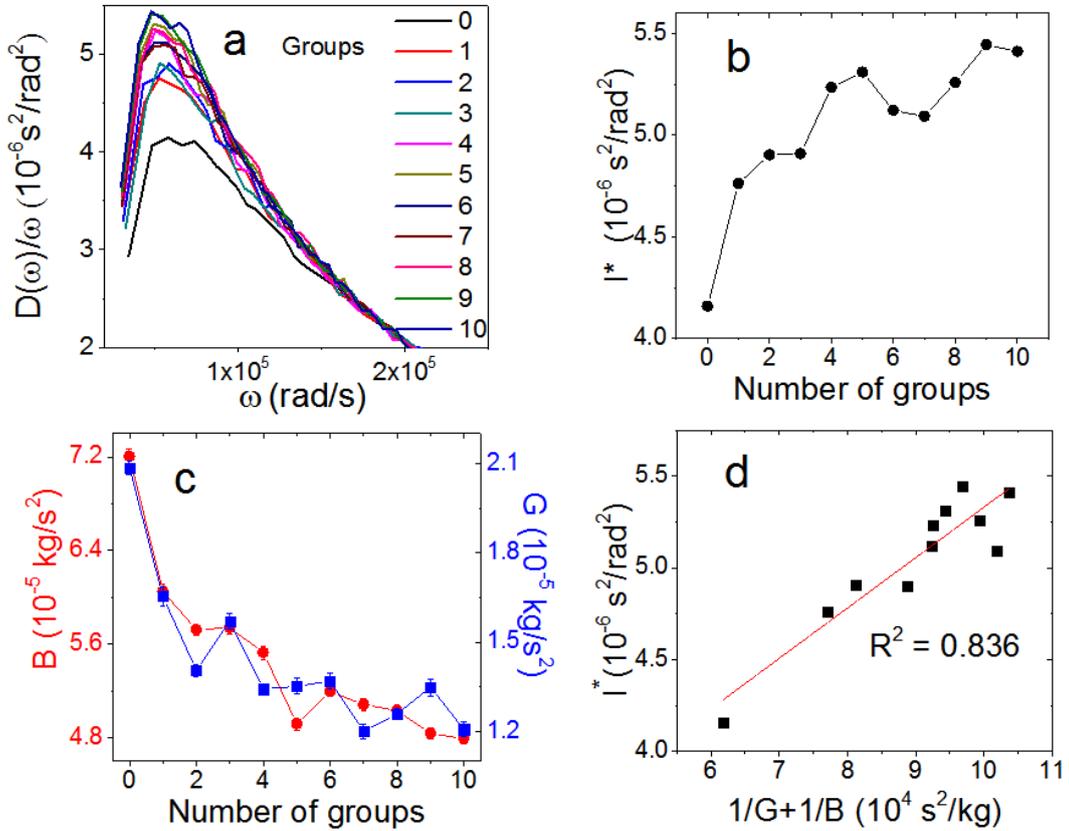

**Figure 2** a, Boson peaks of colloidal samples in different groups of thermal cycles. b, Boson peak intensity (I*) as a function of the number of groups of thermal cycling. c, Evolution of bulk (B) modulus and shear (G) modulus during thermal cycling. d, Simple inverse proportional relation correlating the boson peak intensity to moduli.

### 3.2. Dynamic rejuvenation

Before the displacement covariance matrix method is proposed, accelerated dynamics is a major signature of rejuvenation in colloidal glasses [13-17]. Figure 3a plots the MSD for the samples after 0 to 10 groups of thermal cycles. Each MSD curve reaches its plateau after around 90s. In figure 3b, we summarize the MSD plateau as a function of the number of groups of thermal cycles. Different from the monotonic evolution of boson peak intensity and moduli in figure 2, the MSD plateau fluctuate significantly in different group of thermal cycles. It increases abruptly after the first group of thermal cycling is applied, and almost recovers to its original state with a small value after the second thermal group; the plateau maintains its small value until the seventh groups of thermal cycles. We refer these two groups of thermal cycles after which the system dynamics are significantly accelerated as dynamical rejuvenation DR#1 and DR#2.

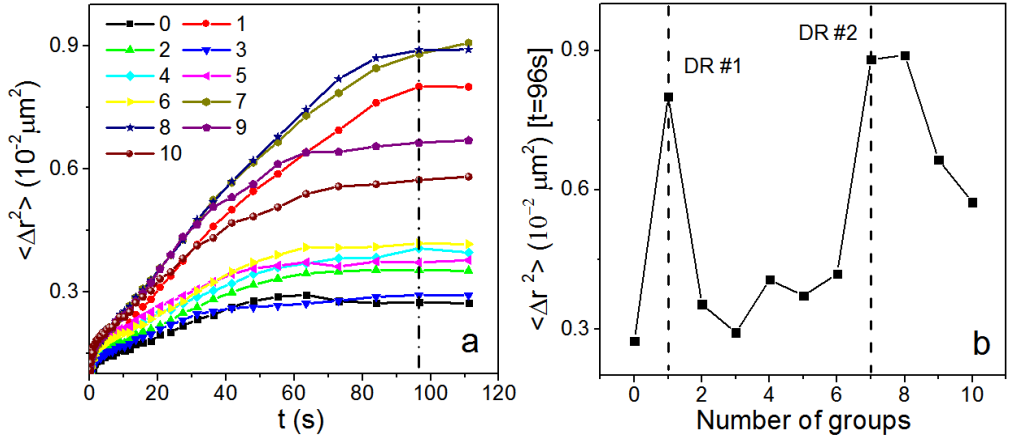

**Figure 3** a, Mean square displacements from different groups of thermal cycles. The dash dot line indicates the time (96s) when we extract the MSD plateau value. b, MSD plateau values as a function of the number of groups of thermal cycles. Dashed lines indicate two dynamic rejuvenation events where the MSD plateau value increases significantly from a group of thermal cycles.

### 3.3. Microscopic origin of thermodynamic rejuvenation

In figure 2 and figure 3, the thermodynamics and dynamics show distinct dependence on the thermal groups. This distinction is suggestive of different microscopic mechanisms underlying the thermodynamic and dynamic rejuvenations in the same samples. Local particle rearrangement, the elementary process of glass deformation and glass relaxation, is believed to be the microscopic culprit of macroscopic rejuvenation [6, 41]. We thus extract the rearrangements induced by thermal cycling and investigate how these rearrangements alter the local properties and hence the macroscopic properties. Local rearrangements are defined as particles that lose their nearest neighbors which are identified by radical Voronoi tessellation [35]. Clusters with less than 5 neighboring particles are ignored [25]. We find that only ~5% of the particles are detected to rearrange during a group of thermal cycles. It is an intriguing question how a small fraction of particle rearrangements lead to significant macroscopic rejuvenation such as ~40% decline of bulk modulus.

We use local stiffness, i.e., the local potential curvature, to characterize a particle's resistance to mechanical deformation. Local stiffness is defined as $K_{ii} =$

$\sqrt{(K_{ii,x}^2 + K_{ii,y}^2)}$, where $K_{ii,x}$ and $K_{ii,y}$ are the diagonal elements in stiffness matrix for particle $i$ [42]. The ensemble-averaged $K_{ii}$ as a function of number of thermal groups is shown in figure 4a and the thermal cycle dependence of $K_{ii}$ is very similar to those of moduli (especially the bulk modulus) in figure 2c, which suggests that $K_{ii}$ is a rough measurement of particle-level modulus. Figure 4b plots the local stiffness change (colored contour) from the first group of thermal cycles as an example of how local rearrangements (white dots) induced by thermal cycling alter the stiffness. Regions surrounding rearrangements tend to suffer from significant stiffness drop and the stiffness decrease could extend to a distance of several particles away from the rearrangement. We quantify the correlation between the rearrangements and the stiffness change by summarizing the average stiffness change from all 10 groups of cycles as a function of the distance to the rearranging particles as shown in the inset of figure 4a. Statistically, rearranging particles suffers from the most significant stiffness decrease with a functioning range over 5 particle diameters, suggesting that local rearrangements induced by thermal cycling are responsible for the softening of the moduli and hence the mechanical rejuvenation.

In order to reproduce the particle-level correlation between boson peak intensity and moduli, we calculate the boson peak intensity for a single particle $D_i(\omega_{\text{bp}})$ and quantify its correlation with $K_{ii}$. $D_i(\omega_{\text{bp}})$ is defined as $D_i(\omega_{\text{bp}}) = \frac{1}{2\Delta\omega} \int_{\omega_{bp}-\Delta\omega}^{\omega_{bp}+\Delta\omega} D_i(\omega') d\omega'$ [40], where $\omega_{\text{bp}}$ is the boson peak frequency, $D_i(\omega')$ is the vibrational density of states for particle $i$ and $2\Delta\omega$ corresponds to thirty vibrational modes around $\omega_{\text{bp}}$. The linear correlation coefficient is calculated to be $C_{A,B} = \frac{\sum_{i=1}^{N}(A_i-<A>)(B_i-<B>)}{\sqrt{\sum_{i=1}^{N}(A_i-<A>)^2}\sqrt{\sum_{i=1}^{N}(B_i-<B>)^2}}$ where A and B correspond to $D_i(\omega_{\text{bp}})$ and $K_{ii}$ and $<.>$ denotes ensemble average. The coefficient is about -0.7 regardless of the number of thermal groups. This strong negative correlation suggests that the boson peak originates from single-particle contribution to the elastic continuum (local stiffness) and justifies the simple inverse relationship between moduli and boson peak intensity in figure 2d at a microscopic level.

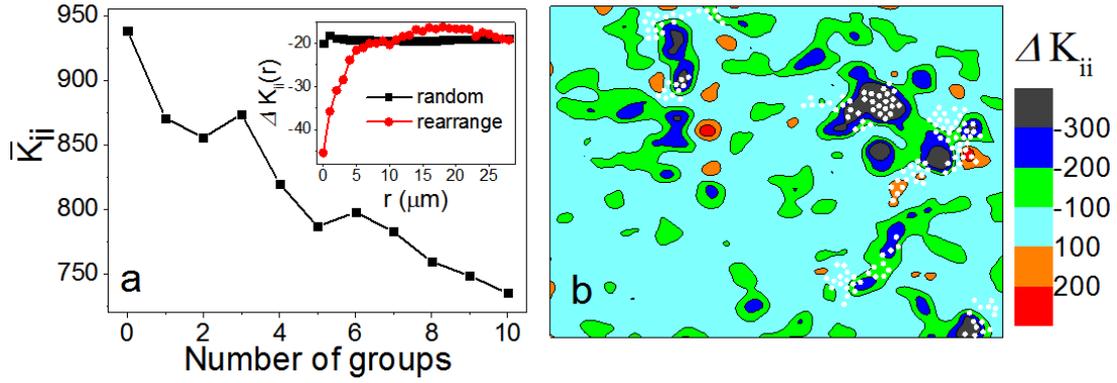

**Figure** 4 a, Ensemble-averaged local stiffness as a function of number of thermal groups. Inset: stiffness change $\Delta K_{ii}$ as a function of the distance to a rearranging particle (red circles) and to a random particle (black squares). b, Distributions of $\Delta K_{ii}$ (contour) and rearranging particles (white dots) induced by a group of thermal cycles.

### 3.4. Microscopic origin of dynamic rejuvenation

In order to investigate into the microscopic mechanisms of dynamic rejuvenation, we calculate the single-particle contribution to MSD, i.e. the local Debye-Waller factor $\alpha_i$. $\alpha_i$ of particle $i$ is defined as $\alpha_i = <[r_i(t) - r_i(0)]^2>_t$, where $r_i(t)$ is the position of particle $i$ at time $t$ and $<.>_t$ denotes the average within the duration where the MSD curve reaches its plateau. Figure 5a-c plots the spatial distribution of $\alpha_i$ before thermal cycling, after the first group of thermal cycles and after the second group of thermal cycles. Before thermal cycling, the $\alpha_i$ of more than 99% particles keeps at values lower than 0.005 μm (figure 5a). The first group of thermal cycles triggers some particles rearranging in the field of view (white dots in figure 5b), followed by localized regions with $\alpha_i$ increasing to values readily above 0.005 μm. The emerging high $\alpha_i$ regions locate right at the position where the largest rearrangements occur. Unlike the microscopic mechanisms of the thermodynamic rejuvenation where a small fraction (~5%) of rearrangements lead to stiffness decline widely spreading in the whole glass sample, the rearrangements create localized fast regions composed of a rather small fraction (10%) of particles by the rearranging regions. The $\alpha_i$ values of the emerging fast regions are extremely higher than the average and lead to the sharp MSD plateau increase in figure 3b even with only 10% of particles facilitated. After the second group of thermal cycles during which new rearrangements occur, the local fast regions disappear with the system MSD plateau recovering to a low value. We confirm that the dramatic MSD plateau jump after the seventh groups of thermal cycles (DR#2) also result from an emergence of localized fast domains. Following previous studies on dynamic heterogeneity [43-47], we cluster particles with 10% highest $\alpha_i$ values as cooperative rearranging regions (CRRs) and emphasize the role of the local fast particles in the system dynamics. CRR clusters with fewer than 5 particles are ignored since separated mobile particles are highly constrained by neighboring less mobile particles and are not likely to dominate the overall dynamics. The total number of clustered particles in CRRs (black squares) is shown in figure 6 and evolves synchronously with the system MSD plateau (figure

3b). This synchronization suggests that the overall dynamics in our system is dominated by localized fast clusters whose emergence would lead to the macroscopic dynamical rejuvenation.

Next, we explore the structural origin of the emergence of CRRs. Local structural entropy $S_2$, a structural parameter highly correlated with particle-level dynamic in binary glasses[35, 48-50], is measured in figure 5 e-f after corresponding thermal groups as in figure 5 a-c. $S_2^i = -\frac{1}{2}\sum_\nu \rho_\nu \int d\boldsymbol{r} \{g_i^{\mu\nu}(\boldsymbol{r})\ln g_i^{\mu\nu}(\boldsymbol{r}) - [g_i^{\mu\nu}(\boldsymbol{r}) - 1]\}$, where $\nu, \mu$ denotes small or big particles, $\rho_\nu$ is the number density of $\nu$ particle, $g_i^{\mu\nu}(\boldsymbol{r})$ is the pair correlation function between particle $i$ of type $\mu$ and the other particles of type ν. In our experiments, the integration is truncated at the third-neighbor shell to avoid the loss of a large fraction of particles due to the boundary effect [35]. $S_2$ is the entropy reduction induced by pair correlation (relative to ideal gas) and high $S_2$ values correspond to less correlated or fragile regions prone to relaxation and deformation in glasses. In figure 5d, before thermal cycling almost all particles have $S_2$ values lower than -17 suggesting that all particles in the well-aged samples are highly correlated with few defects. However, after the first group of thermal cycles as in figure 5e, a high-$S_2$ region underlying rearrangements emerges at the same spot as the high $\alpha_i$ domain (figure 5b). Another group of thermal cycles eliminates the high-$S_2$ region and the high-$\alpha_i$ domain. The correspondence between the high-$\alpha_i$ domain in figure 5 a-c and the high-$S_2$ region in figure 5 d-f qualitatively suggests that the emergence of CRRs could be a result of the activation of high-$S_2$ regions from rearrangements.

We cluster the particles with 10% highest $S_2$ values as structural defects by ignoring clusters with fewer than 5 particles. The number of particles contributing to the structural defects after different groups of thermal cycles is also summarized in figure 6. The size of high-$S_2$ defects and that of CRRs shows a similar dependence on the groups of thermal cycles, although high-$S_2$ particles are less likely to be clustered and the size of high-$S_2$ defects is smaller than that of CRRs. Besides the number synchronization, we confirm the correlation between CRRs and $S_2$ defects by calculating their cluster correlation. First, we define a binarized CRR vector **C** whose number of elements equals the total number of particles $N$. The component element equals unity if the corresponding particle belong to CRRs, otherwise, it equal zero. High-$S_2$ defects are characterized by the other binarized vector **S** with the same length as vector **C**. The elements of **S** are assigned with unity when the corresponding particles belong to the particular $S_2$ defects that have nonzero overlap with **C**. The cluster correlation between CRRs and $S_2$ defects is then quantified by $C_{CRR,defect} = [\frac{\boldsymbol{C}\cdot\boldsymbol{S}}{m} + \frac{(1-\boldsymbol{C})\cdot(1-\boldsymbol{S})}{N-m}]$, where $m$ is the average number of non-zero elements in the two vectors [25, 51]. The cluster correlation averaged from samples after different number of thermal groups is calculated to be 0.6. This high correlation suggests suggest that local high-$S_2$ defects could be the structural origin of CRRs and be responsible for the macroscopic dynamical rejuvenation.

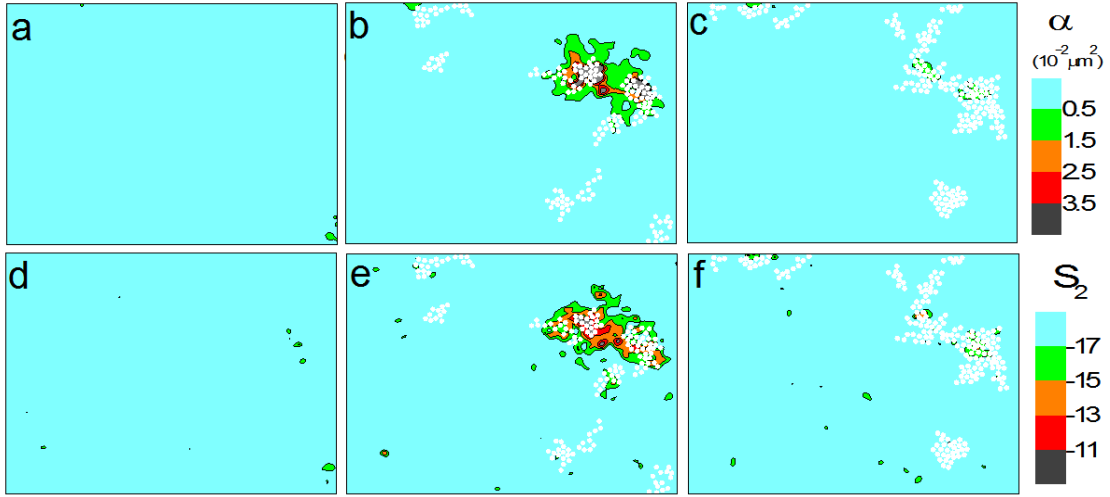

**Figure 5** The spatial distribution of local Debye-Waller factor $\alpha_i$ before thermal cycling (a), after the first group of thermal cycles (b) and after the second group of thermal cycles (c). The spatial distribution of structural entropy $S_2$ before thermal cycling (d), after the first group of thermal cycles (e) and after the second group of thermal cycles (f).

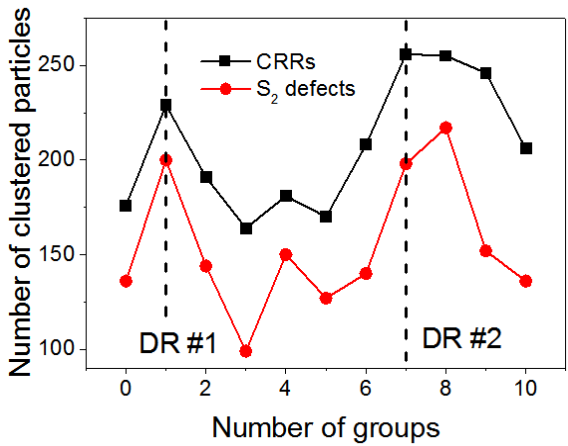

**Figure 6** Number of clustered particles in CRRs (a), and in high $S_2$ defects (b) as a function of number of groups of thermal cycles. Vertical dashed lines indicate the two dynamical rejuvenation processes in figure 3b.

## 4. Conclusion and discussion

Thermal cycling is realized in our colloidal glass systems by optical heating. During thermal cycling, we demonstrate a decoupling between the roughly monotonic thermodynamic rejuvenation and the fluctuated dynamic rejuvenation. Both the thermodynamic and dynamic responses to thermal cycling are induced by a small fraction of local structural rearrangements. However, the same local rearrangements could alter the particle level properties responsible for the thermodynamic and dynamic in a different way. On the one hand, the weakening effects of rearrangements on the elastic continuum spread in the whole sample and lead to the significant thermodynamic rejuvenation. On the other hand, rearrangements change the structural correlation in a localized way, which can either activate or deactivate local cooperative rearranging regions (with high $S_2$ values), leads to a significant mobility change, and results in the fluctuated dynamics.

Our results demonstrate and explain why rejuvenation is not a simple reversal of physical aging. The mechanical and thermodynamic properties are dominated by the elastic continuum, while the dynamics of glasses mainly originate from localized fast regions. Rearrangements weaken the moduli by disrupting the elastic continuum and could influence the local stiffness at a rather long distance to the rearranging regions; whereas dynamic heterogeneity is facilitated only when rearrangements burrow the correlation matrix by creating local defects. The decoupling between macroscopic properties with distinct microscopic mechanisms suggests that glass properties may not be understood within a single unified theoretical frame and the modulation of different glass properties are supposed to follow different principles.

**Acknowledgment**
We thank Chenhong Wang, Rui Liu, Mingcheng Yang, and A. G. Yodh for instructive discussions.